\documentclass[aps,prd,10pt,twocolumn,preprintnumbers,superscriptaddress,showpacs,floatfix,nofootinbib]{revtex4-2}
\usepackage{textcomp,gensymb}
\usepackage{hyperref}
\usepackage{graphicx}
\usepackage{amsfonts,amsmath,amssymb,bm,bbm}
\usepackage{color,soul}
\usepackage[dvipsnames]{xcolor}
\usepackage{slashed}
\usepackage[toc,page]{appendix}
\usepackage{flushend}
\usepackage{physics}
\usepackage{graphicx}
\usepackage{dcolumn}
\usepackage{float}
\usepackage[toc,page]{appendix}
\usepackage{accents}

\hypersetup{
    pdfnewwindow=true,      
    colorlinks=true,       
    linkcolor=blue,          
    citecolor=blue,        
    filecolor=blue,      
    urlcolor=blue        
}

\newcommand\brobor{\smash[b]{\raisebox{0.6\height}{\scalebox{0.5}{\tiny(}}{\mkern-1.5mu\scriptstyle-\mkern-1.5mu}\raisebox{0.6\height}{\scalebox{0.5}{\tiny)}}}}

\begin{document}
\title{New Supernova Constraints on Vector-Mediated Neutrinophilic Dark Sector}

\author{Christopher V. Cappiello}
\email{cappiello@wustl.edu}
\affiliation{Department of Physics and McDonnell Center for the Space Sciences, Washington University, St. Louis, MO 63130, USA}

\author{P. S. Bhupal Dev}
\email{bdev@wustl.edu}
\affiliation{Department of Physics and McDonnell Center for the Space Sciences, Washington University, St. Louis, MO 63130, USA}

\author{Amol V. Patwardhan}
\email{apatwardhan@reed.edu}
\affiliation{Department of Physics, Reed College, Portland, OR 97202, USA}
\affiliation{Department of Physics, New York Institute of Technology, New York, NY 10023, USA}
\affiliation{School of Physics and Astronomy, University of Minnesota, Minneapolis, MN 55455, USA}

\begin{abstract}
Supernova cooling has long been used to constrain physics beyond the Standard Model, typically involving new mediators or dark matter (DM) particles that couple to nucleons or electrons. In this work, we show that the large density of neutrinos inside the neutrinosphere of supernovae also makes them powerful laboratories to study nonstandard neutrino interactions with a {\it neutrinophilic} dark sector, i.e.~DM and mediator particles interacting primarily with neutrinos. In this case, we find that the existing constraints are rather weak, and for a wide range of currently unconstrained parameter space, neutrino annihilation within a supernova could copiously produce such neutrinophilic DM at a large enough rate to cause noticeable anomalous cooling. From the non-observation of such anomalous cooling in SN1987A, we thus set new constraints on neutrino-DM interactions, which provide up to five orders of magnitude improvement in the effective coupling over the existing constraints for DM masses below ${\cal O}$(100 MeV).
\end{abstract}
\maketitle

\section{Introduction}
 
The idea that dark matter (DM) or other exotic particles could be constrained by the observation of core-collapse supernova (CCSN) neutrinos was pointed out~\cite{Ellis:1987pk, Raffelt:1987yt, Turner:1987by} almost as soon as SN 1987A was detected~\cite{Kamiokande-II:1987idp, Bionta:1987qt, Alekseev:1988gp}. If there exist weakly interacting beyond-Standard-Model (BSM) particles, they could be produced in the dense protoneutron star (PNS) and free-stream out of the supernova, altering its thermal evolution and thus affecting the observed neutrino burst. Supernova cooling has since been used to constrain a multitude of BSM particles, including DM candidates, new force carriers, and Kaluza-Klein modes in extradimensional theories~\cite{Raffelt:1996wa, Essig:2013lka}. 
Due to the enormous neutrino densities within supernovae, it is also possible to use them to probe nonstandard physics in the neutrino sector; see e.g. Refs.~\cite{Dicus:1982dk,Kolb:1987qy, Manohar:1987ec, Dicus:1988jh, Fuller:1988ega, Konoplich:1988mj, Kainulainen:1990bn, Shi:1993ee, Kolb:1996pa, Nunokawa:1997ct, Hidaka:2006sg, Hidaka:2007se, Fuller:2008erj, Blennow:2008er, Arguelles:2016uwb, 
Heurtier:2016otg, Das:2017iuj, Dighe:2017sur, Yang:2018yvk,Mastrototaro:2019vug, Shalgar:2019rqe, Suliga:2020vpz, 
Suliga:2020jfa, Akita:2022etk, Chang:2022aas, Chen:2022kal, Fiorillo:2022cdq, Ray:2023gtu,  Manzari:2023gkt,   Fiorillo:2023ytr, Chauhan:2023sci, Carenza:2023old, 
Bhattacharya:2023wzl,
Akita:2023iwq,
Lai:2024mse,Ray:2024jeu, Telalovic:2024cot, Suliga:2024nng}.

In this work, we combine these two ideas, viz., show that neutrino annihilation to DM within a CCSN, followed by the DM free-streaming out of the PNS, could significantly cool the PNS for a wide range of DM masses and couplings. We self-consistently model both the production of DM and its trapping within the neutrinosphere by elastic collisions with neutrinos via a neutrinophilic mediator, in order to compute the extra energy that escapes via the DM from the neutrinosphere. As a result, we get a new stringent constraint on neutrinophilic DM for masses $\lesssim 300$ MeV, for a wide range of couplings consistent with thermal production via neutrino annihilation. This is different from previous studies which considered DM or mediator production in supernovae using couplings to nucleons/electrons/muons~\cite{Fayet:2006sa,
DeRocco:2019jti, Suliga:2020jfa, Croon:2020lrf,Sung:2021swd,Caputo:2021rux,
Cerdeno:2023kqo, Lai:2024mse}, where the existing laboratory constraints are more stringent than our pure neutrinophilic case.

\section{DM-Neutrino Interaction}

For concreteness, we consider a fermionic DM $\chi$ which interacts with the SM neutrinos via a massive vector mediator $Z'$. The relevant part of the interaction Lagrangian in the low-energy effective theory is given by
\begin{equation}
    -\mathcal{L}_{\rm int}\supset g_{\nu} Z'_{\mu}\bar{\nu}\gamma^{\mu}P_L\nu+g_{\chi}Z'_{\mu}\bar{\chi}\gamma^{\mu}\chi\,.
    \label{eq:lag}
\end{equation}
For simplicity, we assume that the $Z'$ couplings to all neutrino flavors are equal. Moreover, we take the $Z'$ to be purely neutrinophilic, i.e., it does not couple to the charged leptons or quarks at the leading order, thus evading most of the laboratory constraints. This can arise, for instance, from a gauge-invariant, ultraviolet (UV)-complete  model that contains a new heavy fermion charged under an extra $U(1)$ gauge symmetry and mixed with the active neutrinos~\cite{Cherry:2014xra, Farzan:2016wym, Farzan:2017xzy, Abdallah:2021npg}. In this case, loop-induced
couplings of $Z'$ to charged leptons and quarks are inevitable~\cite{Chauhan:2020mgv, Chauhan:2022iuh}; however, their effect is subdominant compared to the effect of the tree-level neutrino coupling for the $g_\nu$ values of interest here. Specifically, since the supernova constraint derived here only depends on the product $g_\nu g_\chi$  and the dark-sector coupling $g_\chi$ by itself is very weakly constrained (from DM self-interaction limit), $g_\nu$ can be as small as $10^{-6}$ and still have a sizable effect on supernova cooling, without any other observable consequence. This makes the supernova constraints derived here a powerful new probe of neutrinophilic dark sector physics, irrespective of the UV-completion details, and complementary to the laboratory searches for neutrinophilic $Z'$, e.g. via meson decays~\cite{Laha:2013xua,Bakhti:2017jhm,Bakhti:2018avv,Arcadi:2018xdd, Chauhan:2020mgv, Bahraminasr:2020ssz,Chauhan:2022iuh, Bakhti:2023mvo, Dev:2023rqb, Bai:2024kmt, Dev:2025tdv}, which are only valid for light mediators. 

%%%%%%%%%%%%%%%%%%%%%
\subsection{Relevant Cross Sections}
%%%%%%%%%%%%%%%%%%%%
 
Given the Lagrangian~\eqref{eq:lag}, the DM production inside the supernova happens due to $s$-channel neutrino-antineutrino annihilation via the $Z'$ mediator, with a cross section
\begin{align}
    \sigma_{\nu_\alpha
    \bar{\nu}_\alpha \rightarrow \chi \bar{\chi}} = &\frac{g_{\nu}^2g_{\chi}^2(s + 2m_{\chi}^2)(s-m_\nu^2)}{6 \pi s [(s - m_{Z'}^2)^2 + m_{Z'}^2\Gamma_{Z'}^2]}
    \sqrt{\frac{s-4m_\chi^2}{s-4m_\nu^2}}\,,
    \label{eq:prod}
\end{align}
for each neutrino species $\nu_\alpha$, where $s$ is the squared center-of-momentum (COM) energy, $m_{Z'}$ is the mass of the mediator, $m_{\chi}$ is the mass of the DM\footnote{While the mass of the dark matter is, in principle, modified by forward scattering off the neutrinos in the protoneutron star, we have checked that this correction is negligible in the parameter space we consider.}, $m_\nu$ is the neutrino mass (which can be neglected here, since the neutrinos are relativistic with typical energies of ${\cal O}$(100 MeV) in the supernova core) and $\Gamma_{Z'}$ is the total decay width of the mediator, given by
\begin{equation}
    \Gamma_{Z'} = \frac{g_{\nu}^2 m_{Z'}}{8 \pi} + \frac{g_{\chi}^2 m_{Z'}}{12 \pi}\left(1+\frac{2m_{\chi}^2}{m_{Z'}^2}\right)\sqrt{1-\frac{4m_{\chi}^2}{m_{Z'}^2}}\,.
\end{equation}
Here we have assumed that $Z'$ decay into a DM pair is kinematically allowed, which is always the case in the parameter space we consider. Specifically, we will consider two benchmarks: (i) heavy mediator limit, where $m_{Z'}\gg m_\chi$, and (ii) $m_{Z'}=3m_\chi$, to show how the supernova limits vary with respect to the mediator mass. Eq.~\eqref{eq:prod} agrees with the corresponding equation in Appendix B of Ref.~\cite{Manzari:2023gkt}. Here, we use a pre-factor of 1 (corresponding to only left-helicity of neutrinos) instead of 1/4 for averaging over initial states. 

The same interaction Lagrangian~\eqref{eq:lag} also leads to DM annihilation back into neutrinos, with the corresponding cross section for each neutrino flavor $\alpha$ (assuming a flavor-universal coupling) given by 

\begin{align}
    \sigma_{\chi \bar{\chi} \to \nu_\alpha\bar\nu_\alpha} = &\frac{g_{\nu}^2g_{\chi}^2(s + 2m_{\chi}^2)(s - m_\nu^2)}{24 \pi s \, [(s - m_{Z'}^2)^2 + m_{Z'}^2\Gamma_{Z'}^2]}
    \sqrt{\frac{s-4m_\nu^2}{s-4m_\chi^2}}\,.
    \label{eq:ann}
\end{align}
This agrees with the form which appears in Appendix B of Ref.~\cite{Manzari:2023gkt}  (with $V_l = - A_l = 1/2$), and also reduces to Eq.~(E.8a) of Ref.~\cite{Dev:2025tdv} in the nonrelativistic limit (with $g_{\chi L} = g_{\chi R} = g_\chi$). Here, a prefactor of 1/4 was used while taking average over initial states (1/2 each for $\chi$ and $\bar\chi$).

The DM particles, once produced, can also scatter with the ambient neutrinos via the same $Z'$ interaction~\eqref{eq:lag}. The differential scattering cross-section for $\overset{\brobor}\chi \, \overset{\brobor}\nu \to \overset{\brobor}\chi \, \overset{\brobor}\nu$ can in general be written in a frame-independent form as 
\begin{equation}
    \frac{\textrm{d}\sigma}{\textrm{d}t} = \frac{\overline{|\mathcal{M}|^2}}{64\pi s} \frac{1}{p_\text{cm}^2}\, ,
\end{equation}
where $\overline{|\mathcal{M}|^2}$ is the appropriately spin-averaged squared matrix element for the scattering process, and $p_\text{cm}$ is the magnitude of the 3-momentum of one of the incoming particles in the COM frame. For the model Lagrangian~\eqref{eq:lag}, the full expression is given by 
\begin{equation}
    \frac{\textrm{d}\sigma}{\textrm{d}t} \bigg|_{\chi\nu_\alpha\to\chi\nu_\alpha} = \frac{g_\chi^2 g_\nu^2}{8 \pi (s-m_\chi^2)^2} \frac{2 (s-m_\chi^2)^2+2 s t+t^2}{(t-m_{Z'}^2)^2} \, ,
    \label{eq:scattering}
\end{equation}
where we have used $p_\text{cm} = (s-m_\chi^2)/2\sqrt{s}$ for $\chi\nu \to \chi\nu$. In this model, the cross-sections for $\chi\bar\nu$, $\bar\chi\nu$ and $\bar\chi\bar\nu$ are the same as above.

\section{Input Parameters}

\begin{figure}[b!]
    \centering
    \includegraphics[width=\linewidth]{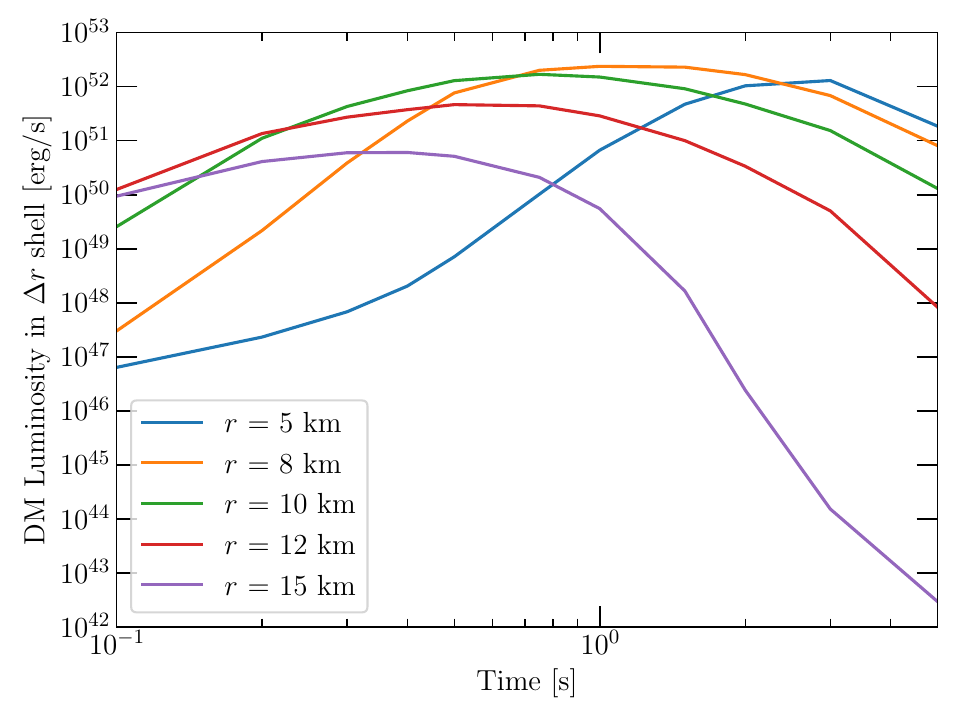}
    \caption{DM luminosity (i.e.~emissivity,  multiplied by the volume of a spherical shell with radius $\Delta r$), as a function of time at several different radii. Here $m_{\chi} = 1$ MeV, $m_{Z'}=10$ GeV, $g_{\nu}g_{\chi} = 10^{-6}$, and $\Delta r = 1$ km. }
    \label{fig:emissionvstime}
\end{figure}
To model the properties of the neutrino population, we use the output of a 1-D CCSN simulation for a 20 $M_{\odot}$ progenitor PNS~\cite{jankasim} (see also Ref.~\cite{Mirizzi:2015eza} for a description of the physics underlying the simulation). This simulation gives profiles of density, temperature and chemical potential both as a function of radius and for several points in time post core bounce. Through the radial dependence of the temperature and chemical potential, $\dot{E}$ is implicitly a function of radius, as will be made explicit below. The details of the simulation and some illustrative plots are given in Appendix~\ref{Appendix A}.

Figure~\ref{fig:emissionvstime} shows the amount of neutrino energy going into DM production in several spherical shells as a function of time post bounce. Here the couplings are chosen to be small enough that DM free-streams out of the neutrinosphere without being trapped (see below). We see that the highest emission is at about 8 km, and peaks at around 1 second post-bounce. For this reason, we use simulation results for 1 second post-bounce to compute our limits, as this is when the DM emission should be maximized. Note also that 8 km, the radius with the highest emission, is also where the temperature is maximized at 1 second post-bounce (see Appendix~\ref{Appendix A}).

%%%%%%%%%%%%%
\section{Emissivity Calculation}
%%%%%%%%%%%%%

Assuming spherical symmetry, at a given radial distance $r$ from the center of the supernova core, the rate of neutrino energy going into DM production is given by
\begin{equation}
\begin{split}
    \dot{E} = 
    \frac{8\pi^2}{(2\pi)^6}\int \textrm{d}E_1 \textrm{d}E_2 \textrm{d}\cos{\theta} \:f_1 f_2 E_1^2 E_2^2 (E_1 + E_2) \\ 
    \times \, \sigma_{\nu \bar{\nu} \rightarrow \chi \bar{\chi}} v_{\rm rel}\,,
    \label{eq:emissivity}
\end{split}
\end{equation}
where $v_{\rm rel}\simeq \sqrt{2(1-\cos\theta)}$  is the relative velocity between the annihilating neutrino and antineutrino, and the factors $f_{1,2}$ are the Fermi-Dirac distributions of the neutrinos and antineutrinos, given by
\begin{equation}
    f_i = \frac{1}{1 + e^{[E_i \pm \mu(r)]/T(r)}}\,,
\end{equation}
where the minus (plus) sign corresponds to neutrinos (antineutrinos). In what follows, we suppress the radial dependence, with the understanding that the temperature, chemical potential and density of each neutrino species, are all implicitly functions of the position.

The angular dependence enters the calculation in two ways: through the relative velocity $v_{\rm rel}$ and through the squared COM energy $s = 2E_1E_2(1-\cos{\theta})$. Writing the angular dependence explicitly, we perform the angular integral in Eq.~\eqref{eq:emissivity} analytically and arrive at the emissivity formula
\begin{align}
   & \dot{E} =  \frac{1}{8\pi^4}\int \textrm{d}E_1 \textrm{d}E_2 f_1 f_2 E_1^2 E_2^2 (E_1 + E_2) \, \nonumber \\
   & \qquad \times \frac{g_{\nu}^2g_{\chi}^2}{16 \pi [m_{Z'}^4 + m_{Z'}^2\Gamma_{Z'}^2]} I_c(E_1,E_2,m_{\chi})\,, 
    \label{eq:emissivity1} \\
    {\rm where}\quad & I_c =  \frac{16}{5}(2E_1E_2 + 3m_{\chi}^2)\left(1-\frac{m_{\chi}^2}{E_1E_2}\right)^{3/2}. \nonumber 
\end{align}

\subsection{DM Trapping}
 
It is not enough to merely produce the DM particles: they must also escape from the neutrinosphere. For small enough couplings, virtually all of the DM will escape without scattering. But if the coupling is made large, a DM particle may be trapped, or at least redistribute energy across the neutrinosphere before escaping. Here we restrict ourselves to DM that escapes without scattering.

Because the neutrinos are traveling relativistically, and can be highly degenerate, the mean free path cannot be computed as simply $1/(n_{\nu} \sigma_{\nu\chi\to \nu\chi})$, where $n_\nu$ is the neutrino number density.  First, the neutrino-DM scattering rate is suppressed if the outgoing neutrino energy $E_{\rm out}$ is Pauli-blocked, which is accounted for by the replacement
\begin{equation}
    \frac{\textrm{d}\sigma}{\textrm{d}t} \rightarrow [1-f_i(E_{\rm out},\mu,T)]\frac{\textrm{d}\sigma}{\textrm{d}t}\,,
\end{equation}
where $t=-2E^2_\nu(1-\cos\theta)$ is the Mandelstam variable, with $E_\nu$ defined in the COM frame. 
We can thus define a ``Pauli-blocked cross section" that takes Pauli-blocking into account:
\begin{equation}
    \sigma_{\textrm{PB}} = \int [1-f_i(E_{\rm out},\mu,T)]\frac{\textrm{d}\sigma}{\textrm{d}t} \textrm{d}t\,.
\end{equation}
Second, the fact that the neutrinos are not stationary must be accounted for. In terms of the DM velocity $\beta$ and the angle between DM and neutrino velocities $\theta$, we can define the relative velocity factor
\begin{equation}
    v_{\rm rel}(E_\chi,\theta) = \sqrt{1 - 2\beta \cos\theta + \beta^2}\,.
\end{equation}
With both of these factors accounted for, the scattering rate for a DM particle with neutrinos of species $\alpha$ is
\begin{equation}
\Gamma_i(E_\chi,\theta,\phi) = \frac{n_{\nu_{\alpha}} v_{\rm rel}(E_\chi,\theta)}{4\pi  [-2T^3{\rm Li}_3(-e^{\mu/T})]} \int \textrm{d}E_{\nu}E_{\nu}^2f(E_{\nu}) \,\sigma_{\textrm{PB}}\, \,.
\end{equation}
Here the factor $-2T^3{\rm Li}_3(x)$, with Li$_3(x)$ being the polylogarithm function, is a normalization factor to the Fermi-Dirac distribution times density of states, the function which determines the probability of a given neutrino energy.

We thus define the inverse mean free path, reintroducing the explicit position dependence:

\begin{equation}
    \lambda^{-1}(E_\chi,\textbf{r}) = \frac{2\pi}{\beta}\int \sum_i \Gamma_i(E_\chi,\theta,\phi)\, \textrm{d}\cos\theta\,.
\end{equation}
From this, we can compute the probability for a DM particle to escape the neutrinosphere without scattering:
\begin{equation}
    P_0(E_\chi,r) = \textrm{exp} \left(-\int \frac{\textrm{d}l}{\lambda(E_\chi,\textbf{r}')}\right)\,,
    \label{eq:prob}
\end{equation}
where $l$ is the path traveled by the DM, which in general is a function of $r$, as well as the angular variables $\theta$ and $\phi$.  If we imagine that after being produced at a radius $r_0$, all the DM particles travel on radial trajectories directly out of the neutrinosphere, Eq.~\eqref{eq:prob} simplifies to
\begin{equation}
    P_0(E_\chi,r_0) = \textrm{exp} \left(-\int_{r_0}^{\infty} \frac{\textrm{d}r}{\lambda(E_\chi,r)}\right)\,.
    \label{eq:prob2}
\end{equation}
Although assuming a radial trajectory underestimates the true value of the integral in the exponent for most DM particles, it is accurate to within a factor of 2 for a substantial fraction of the DM, depending on the value of $r_0$. It would also be overly conservative to assume that only particles that escape without scattering can result in cooling: DM particles that scatter once or a couple times could still carry away significant energy. Therefore, we consider the above formula a reasonable estimate of the suppression of the energy emission due to trapping.

\begin{figure}[t]
    \centering
    \includegraphics[width=\linewidth]{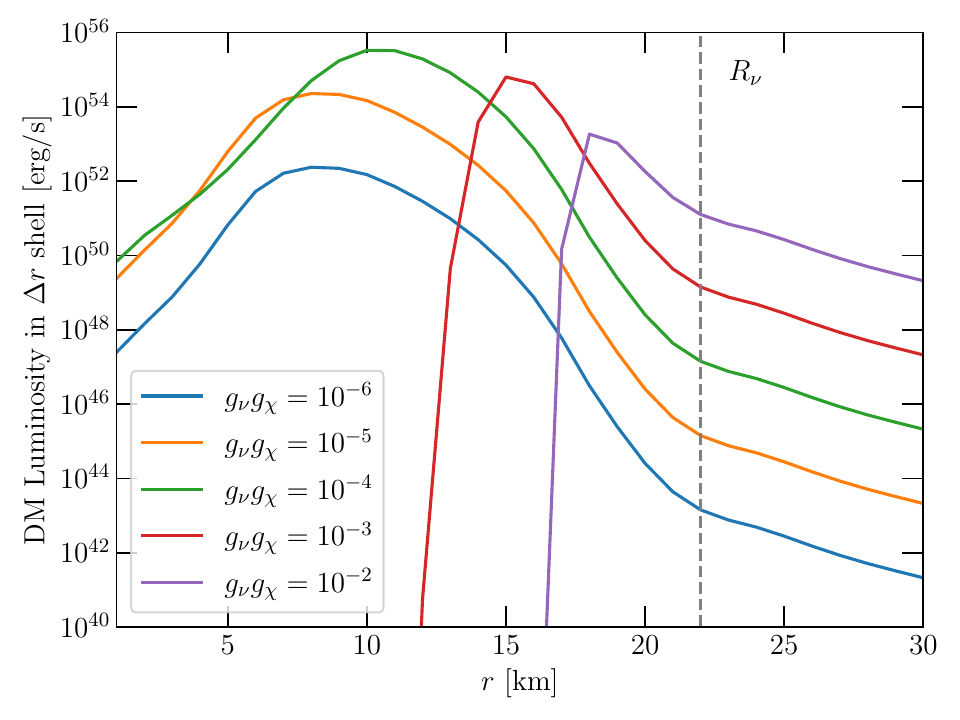}
    \caption{DM luminosity (including trapping effect) at $\textrm{t} = 1$ s in each 1-km-wide spherical shell as a function of radius. Here $m_{\chi} = 1 $ MeV, and $m_{Z'} = 10 $ GeV. Note that the horizontal axis starts at 1 km, because the value on the vertical axis goes to zero at $r=0$. }
    \label{fig:emissionprofile}
\end{figure}

\begin{figure*}[t]
    \centering
    \includegraphics[width=0.49\linewidth]{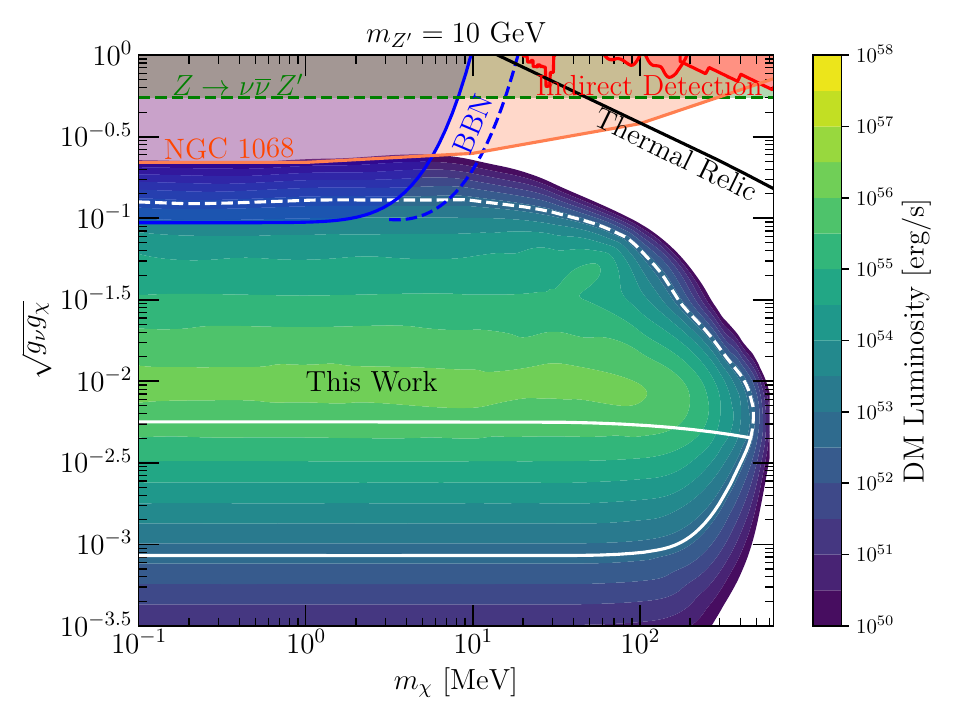}
    \includegraphics[width=0.49\linewidth]{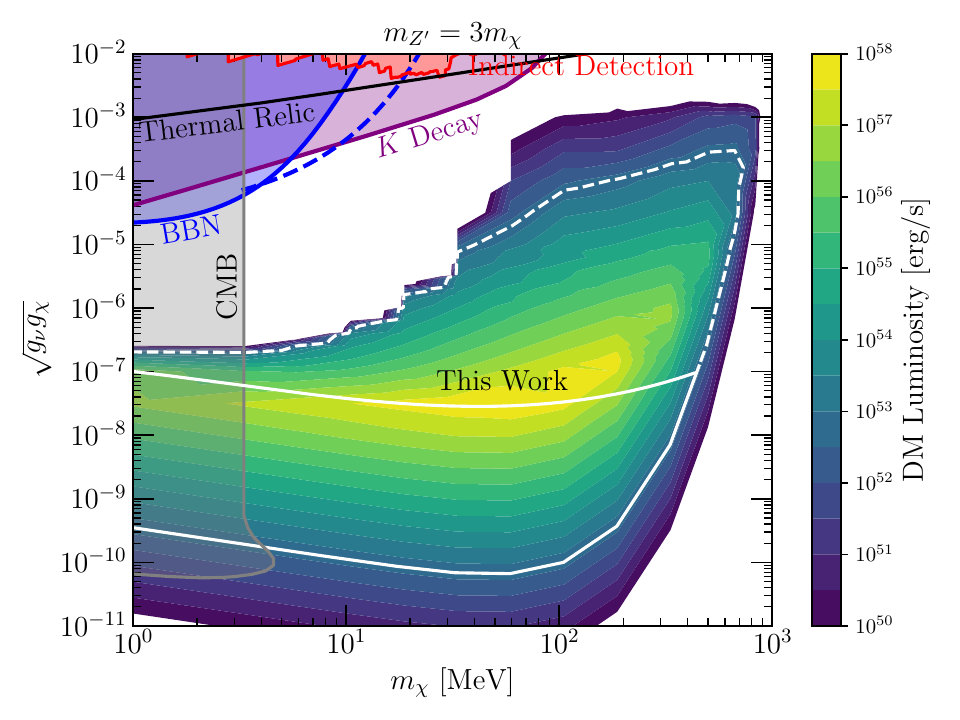}
    \caption{DM luminosity as a function of the DM mass $m_\chi$ and the product of the mediator couplings to neutrinos $(g_\nu)$ and DM  ($g_{\chi}$), for a vector mediator with mass $m_{Z'} = 10$~GeV (left) and $m_{Z'} = 3m_{\chi}$ (right). The white contour is our SN1987A exclusion limit, set where the emissivity equals $5 \times 10^{52}$ erg/s. The dashed upper curve shows the maximum coupling that can be probed using the approach described in this work; the solid white line running through the shaded region shows the coupling above which trapping becomes significant (see text for details). For comparison, we also show exclusion regions from BBN (shaded and dashed blue; see text for details), DM indirect detection searches from Borexino, KamLAND, and Super-K data  (red)~\cite{Olivares-DelCampo:2017feq, Arguelles:2019ouk}, neutrino attenuation in DM spikes (orange)~\cite{Cline:2023tkp}, kaon decay (purple)~\cite{Dev:2025tdv}, and invisible $Z$ decays (dashed green; assumes $g_{\nu} = g_{\chi}$)~\cite{Laha:2013xua, Dev:2024ygx}. The black line corresponds to the thermal DM relic density. }
    \label{fig:limitplot}
\end{figure*}

After being produced, DM can also annihilate back into neutrinos inside the neutrinosphere. Whether this effect is significant depends crucially on the mass of the mediator and of the DM. For a heavy mediator, this effect can safely be neglected: when the mediator mass is much larger than the energies of the neutrinos and produced DM, we can make the replacements $(s - m_{Z'}^2)^2 \rightarrow m_{Z'}^4$ and $(t - m_{Z'}^2)^2 \rightarrow m_{Z'}^4$ in Eqs.~\eqref{eq:prod}, \eqref{eq:ann}, and \eqref{eq:scattering}, so the scattering and annihilation cross sections are comparable at relativistic energies. Even for the largest DM luminosity that can be produced (which requires $g_{\nu}g_{\chi} \simeq 10^{-4}$; see Fig.~\ref{fig:limitplot}, left), we have checked that the ambient neutrino number density exceeds the produced DM number density by about an order of magnitude. For larger couplings, DM and neutrinos deep inside the neutrinosphere may come into thermal equilibrium. But at those couplings, the DM luminosity is dominated by production at larger radii (see Fig.~\ref{fig:emissionprofile}), where the produced DM density is still much lower than the neutrino density. Therefore, we do not include the DM annihilation in the escape probability for the case of a heavy mediator.

However, when we consider a light mediator, the replacement $(s - m_{Z'}^2)^2 \rightarrow m_{Z'}^4$ is no longer valid. In particular, the resonance in Eq.~\eqref{eq:ann} can no longer be neglected. This resonance enhances the DM self-annihilation cross section relative to the scattering cross section with neutrinos, so the self-annihilation process must be accounted for as well. Following Ref.~\cite{Manzari:2023gkt}, we use the narrow-width approximation to simplify the calculation of the annihilation rate. We assume that the density of DM is equal to the volumetric DM production rate divided by the time it takes DM to escape the neutrinosphere. We compute the inverse mean free path for annihilation and add it to the inverse mean free path due to scattering with neutrinos to compute the total inverse mean free path, and use Eq.~\eqref{eq:prob2} to compute the probability of escaping just as before.

Using the average DM energy $E_{\rm avg}$ to compute the escape probability, we define the emissivity after trapping is taken into account as
\begin{equation}
    \mathcal{E}(r) = P_0(E_{\rm avg},r)\dot{E}(r)\,,
\end{equation}
with $\dot{E}$ given by Eq.~\eqref{eq:emissivity1}. In Fig.~\ref{fig:emissionprofile}, we show the DM luminosity per spherical shell for a range of different coupling values. For concreteness, we choose a DM mass of 1 MeV and a mediator mass of 10 GeV. When the couplings are small, the luminosity peaks at about 8 km (where the temperature peaks), and increasing the couplings increases the total energy emission. However, when the couplings become large, any DM produced near the center of the neutrinosphere is effectively trapped, and the emission is dominated by larger radii. Integrating these emission profiles over $r$ yields the total luminosity for a given choice of model parameters, namely, couplings and particle masses.

\section{Results}
\subsection{Heavy Mediator}

Figure~\ref{fig:limitplot} shows the total energy loss rate to DM at 1 second post-bounce for a range of model parameters. In the left panel, we set the mediator mass $m_{Z'} = 10$ GeV, and scan over values of DM mass and couplings. As both the emission and trapping depend only on the product $g_{\nu}g_{\chi}$, we do not vary the couplings independently, instead scanning over values of their product. For small couplings ($g_{\nu}g_{\chi} \lesssim 10^{-6}$), the DM production rate is too small to have a noticeable effect on the supernova. For large couplings ($g_{\nu}g_{\chi} \gtrsim 10^{-2}$), the production rate is large, but the DM is trapped effectively enough that very little energy escapes. Such trapped DM could still redistribute energy within the neutrinosphere, potentially leading to more subtle but still detectable signals, but we do not consider this effect here, as it requires a detailed supernova simulation which is beyond the scope of this work. At the right edge of the plot, the DM is too heavy to be efficiently produced (note that the typical neutrino energies are $\mathcal{O}$(100) MeV; see the plots in Appendix~\ref{Appendix A}).

In the central, multicolored region of Fig.~\ref{fig:limitplot}, the DM luminosity may be substantial compared to the neutrino luminosity. We set our limit by requiring that the luminosity not exceed $5 \times 10^{52}$ erg/s, which is approximately $10\%$ of the maximum neutrino luminosity produced by the simulations of Ref.~\cite{jankasim}. This also agrees roughly with the ``Raffelt criterion" based on observations of SN 1987A, namely that any exotic energy loss should not exceed the neutrino luminosity of $\approx (3-5) \times 10^{52}$ erg/s~\cite{Raffelt:1990yz}. Within the solid white contour, the DM emission would be substantial enough to alter the evolution of the supernova, and additionally, the optical depth for dark matter traveling from the point of peak dark matter production (a radius of 8 km) to the edge of the neutrino density is less than 1; within this region, trapping is largely negligible. Within the dashed contour, trapping becomes important, but by modeling trapping using the approach described in the previous Section, we still find this parameter space to be ruled out based on the same energy-loss criterion. We also see that only slightly within this contour, the emission can be an order of magnitude higher, such that requiring a somewhat larger luminosity of, e.g., $10^{53}$ erg/s would not substantially change our limit. Similarly, we do not expect the current modeling uncertainties on the CCSN parameters~\cite{Janka:2025tvf} to significantly change our results either, as shown in case of other BSM particles~\cite{Chang:2018rso, Balaji:2022noj}.

The black line (in the top right corner of Fig.~\ref{fig:limitplot} left panel) denotes the parameters required for the DM to obtain the correct relic abundance via thermal freeze-out. This is obtained by using Eq.~\eqref{eq:ann} in the non-relativistic limit, which gives 
\begin{equation}
    \langle\sigma_{\chi\bar\chi\to \nu\bar\nu} v\rangle \simeq \frac{g_{\nu}^2g_{\chi}^2}{2\pi}\frac{m_{\chi}^2}{(4m_{\chi}^2 - m_{Z'}^2)^2 + m_{Z'}^2 \Gamma_{Z'}^2} \, ,
\end{equation}
that is then set to the required value of $\approx (2-5)\times 10^{-26}~{\rm cm}^3/{\rm s}$, depending on the DM mass. We use the exact numerical solution from Ref.~\cite{Steigman:2012nb} to obtain our relic density constraint in Fig.~\ref{fig:limitplot}.
Although the parameter space we constrain does not quite reach the thermal relic line, 
it effectively constrains various other possibilities of obtaining the correct relic abundance, such as freeze-in~\cite{Hall:2009bx}, DM annihilation to lighter dark states~\cite{Bhattiprolu:2023akk}, or entropy injection into the SM plasma following DM decoupling~\cite{Evans:2019jcs}.

In Fig.~\ref{fig:limitplot}, the parameter space in blue is excluded by big bang nucleosynthesis (BBN) considerations. This limit is derived by setting the Hubble rate $H$ equal to the DM production rate [cf.~Eq.~\eqref{eq:emissivity} without the $(E_1+E_2)$ factor, divided by the neutrino density] at $T=1$ MeV, so that the DM is not in equilibrium with the thermal plasma at the time of neutrino decoupling. Note that the constraint becomes flat at low $m_{\chi}$ because, in this limit, $m_{\chi}^2 < s$ and $\sigma_{\nu \bar{\nu} \rightarrow \chi \bar{\chi}}$ becomes independent of mass. Another BBN limit can also be set by comparing the rate for the opposite process $\chi\bar\chi\to \nu\bar\nu$, i.e. $\Gamma_\chi=n_\chi \langle \sigma_{\chi\bar\chi\to \nu\bar\nu}v_\chi\rangle$, to the Hubble rate, because an excess neutrino production could alter the neutron-to-proton ratio at the time of BBN, or delay the neutrino decoupling. However, setting such a limit requires an assumed DM density. Although the parameter space we show lies largely below the thermal relic line, for reference we compute the BBN bound that results from a thermal DM density of $ n_\chi =g \left(m_{\chi}T/2\pi\right)^{3/2} e^{-m_{\chi}/T}$ 
(with $g=2$ for DM),  
resulting in the dashed blue curve in Fig.~\ref{fig:limitplot}, which extends to slightly higher masses. We stop this curve at $m_{\chi} = 3$ MeV because the formula we use for the DM density assumes that the DM is nonrelativistic.  

The pink region is excluded based on the non-observation of attenuation of high energy neutrinos produced in the nearby active galaxy NGC 1068~\cite{IceCube:2022der}, assuming that it contains a DM spike~\cite{Cline:2023tkp}. There exist additional constraints based on scattering between spiked DM and neutrinos from the blazar TXS 0506+05~\cite{Cline:2022qld, Ferrer:2022kei} and  tidal disruption events (TDE) such as  AT2019dsg~\cite{Fujiwara:2023lsv}, which are weaker than the NGC 1068 bound in the parameter space considered here, and hence, not shown in Fig.~\ref{fig:limitplot}.  Similarly, there exist cosmological constraints from small-scale structure formation and galactic density profiles~\cite{Akita:2023yga,Heston:2024ljf}, as well as from cosmic microwave background (CMB) and large-scale structure~\cite{Mangano:2006mp, Wilkinson:2014ksa}, which are however weaker for heavy mediators and do not show up in the parameter space of our interest. 

There also exist constraints on our parameter space from DM indirect detection limits. We show limits from 3 analyses of Super-K data, two of which were first presented in Ref.~\cite{Arguelles:2019ouk} and the third is reproduced from Ref.~\cite{Olivares-DelCampo:2017feq}, collectively shown as the red-shaded region.

The model we consider also allows for neutrino self-interactions via the $Z'$ mediator. We show the constraint on this interaction arising from invisible $Z$ decays~\cite{Laha:2013xua, Dev:2024ygx}. As this limit is independent of any DM interaction, it does not actually constrain $g_{\chi}$; for reference, we plot this bound under the assumption that $g_{\chi} = g_{\nu}$. Note that even large neutrino self-interactions have little impact on the supernova dynamics~\cite{Fiorillo:2023ytr,Fiorillo:2023cas}, which justifies our use of the standard CCSN simulation results~\cite{jankasim} for our purpose.

Our model also allows for DM self interactions, but self interaction bounds from e.g. the Bullet Cluster~\cite{Markevitch:2003at} are not strong enough to constrain the parameter space shown here. 

\subsection{Light Mediator}

The DM luminosity for the light mediator case has features similar to those shown in Figs.~\ref{fig:emissionvstime} and \ref{fig:emissionprofile}. But the production and trapping rates can be very different. Our results for the case where $m_{Z'} = 3m_{\chi}$ are shown in Fig.~\ref{fig:limitplot}  right panel. Our constraint reaches much smaller couplings than in the case of a heavy mediator, not simply because of the $1/m_{Z'}^4$ scaling of the production cross section in the heavy-mediator limit, but also because of the resonant enhancement present in this case. On the other hand, the corresponding resonance in the DM self-annihilation cross section prevents our constraint from reaching maximum couplings as large as in the heavy-mediator case, because DM efficiently self-annihilates when the coupling becomes too large. Crucially, for a light enough mediator, this process becomes much more important than DM scattering with neutrinos.

In addition to our limit itself, we also show the emissivity in multicolored shading. For DM masses much below 100 MeV, we see that the emissivity drops much more sharply at large couplings than it does at larger masses. The reason for this is the transition from trapping via scattering with neutrinos (at large mass) to trapping via self-annihilation (at low masses). Because the probability of a DM particle annihilating before it escapes depends on the production rate of DM itself, the annihilation rate increases more rapidly with increasing couplings than the probability of scattering does.

In the case of a light mediator, there exist constraints not only on the DM itself, but also on the mediator, as the mediator can be produced on-shell and then decay into $\nu\bar{\nu}$, thus changing $\Delta N_{\rm eff}$. One such constraint on light mediators coupled to neutrinos based on the CMB data can be found in Ref.~\cite{Li:2023puz}, and is shown by the gray region in Fig.~\ref{fig:limitplot}  right panel. An additional bound from kaon decay~\cite{Dev:2025tdv} is shown in purple. These bounds are shown in addition to the other bounds discussed above in the case of a heavy mediator.\footnote{A recent paper~\cite{Farzan:2026wek} suggests that the NGC 1068 bound is much weaker than that shown in Ref.~\cite{Cline:2023tkp} in the light mediator limit.}

%%%%%%%%%%%%%%%%%
\section{Discussion and Conclusion}
%%%%%%%%%%%%%%%%%%
Here, we considered fermion DM and vector mediator for concreteness, but it is straightforward to generalize our work to any combination of DM and mediator type (scalar, pseudo-scalar, fermion, vector and axial-vector).

We reported limits under two assumptions for the mediator mass: first, we considered a ``heavy'' mediator with a fixed mass of 10 GeV, and second, we considered a ``light" mediator with mass equal to three times the DM mass. For mediators heavier than a few GeV, our heavy-mediator results can easily be rescaled by simply keeping the ratio $g_{\nu}^2g_{\chi}^2/m_{Z'}^4$ constant. For lighter mediators, the rescaling is less straightforward. Although the typical neutrino energies are of order 100 MeV, even for mediators as heavy as 1 GeV, the resonance at $s = m_{Z'}^2$ can be reached at a high enough rate to noticeably affect the total neutrino annihilation rate. The resulting limits are much stronger for a light mediator than would be suggested by the simple rescaling above, due to the presence of this resonance. At the same time, resonant reannihilation of DM into neutrinos is significant in this case, again due to the resonance in the self-annihilation cross section.

The results presented here are similar to those presented in Ref.~\cite{Manzari:2023gkt}, \textcolor{blue}{but} with several important distinctions. First, the result of Ref.~\cite{Manzari:2023gkt} for the commonly used benchmark $m_{\chi} = m_{Z'}/3$ is presented only for the full $L_{\mu} - L_{\tau}$ model, meaning that constraints are not available for a neutrinophilic model in this case. Our treatment of attenuation is also significantly different. In presenting our results, we have also compiled a wide range of constraints on neutrinophilic models, and in the process have derived a new BBN constraint based on dark matter annihilation to neutrinos.

Similarly, supernova cooling has also been used to set limits on heavy neutral leptons, which can be produced through neutrino annihilation, as considered in Ref.~\cite{Carenza:2023old}.  Here, the production rate receives additional contributions from charged leptons as well as nucleons. Moreover, the heavy neutral leptons being considered there are also unstable, while our dark matter particles are stable, meaning that decay of the dark matter within the supernova is not a concern. Nevertheless, we have verified that the order of magnitude of the cooling bounds obtained here are similar to those in the heavy neutral lepton case neglecting the nucleon production channel.

In conclusion, we have shown that the non-observation of anomalous cooling in SN1987A sets new stringent constraints on neutrino-DM interactions, providing up to three (five) orders of magnitude improvement over the existing constraints for DM masses below ${\cal O}$(100 MeV) for heavy (light) mediators. Although we have only considered vector mediators in this analysis, we expect the results to be similar in nature for scalar mediators with interactions of the type $\bar{\nu}\nu\phi$ and $\bar{\chi}\chi\phi$. On the other hand, interactions of the type $\bar{\nu}\chi\phi$ are more involved and will be the subject of a follow-up paper.

\begin{figure*}[!t]
    \centering
    \includegraphics[width=0.49\textwidth]{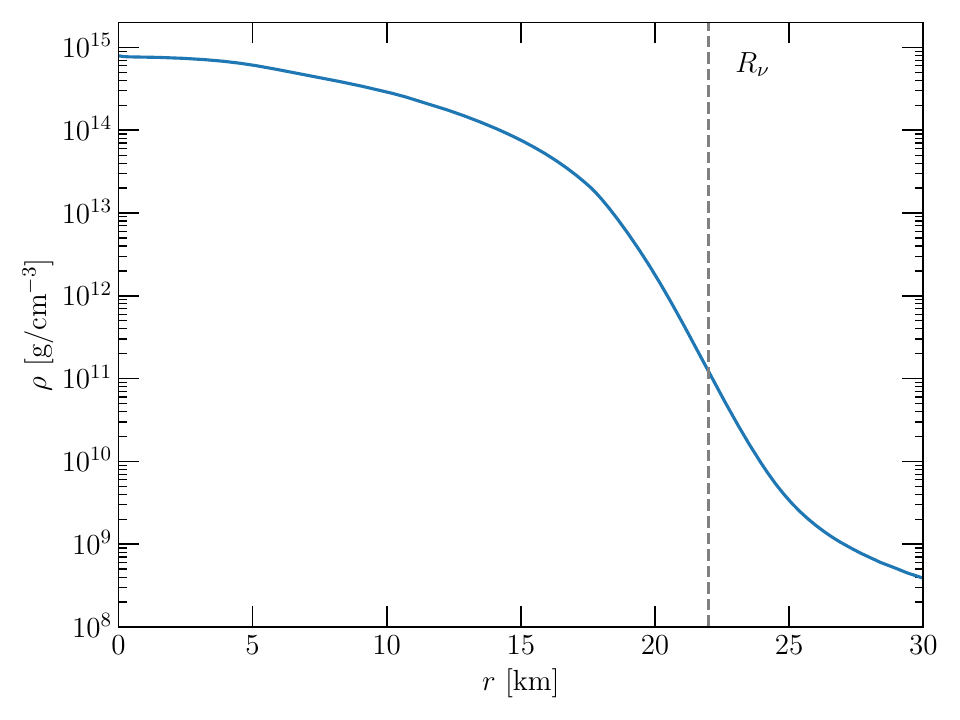}
    \includegraphics[width=0.49\textwidth]{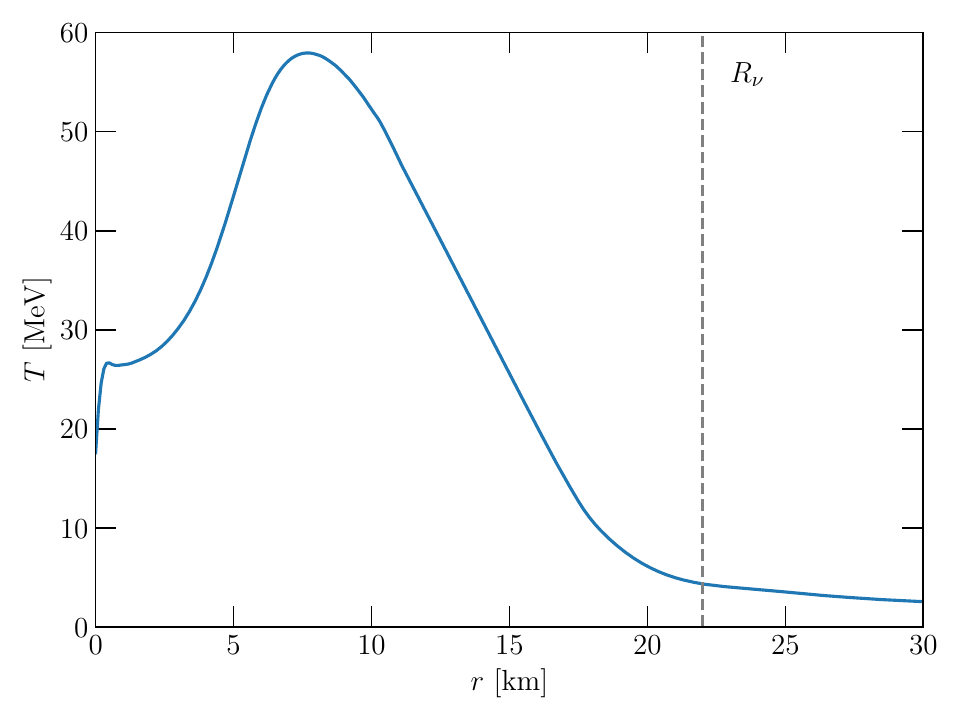}
    \includegraphics[width=0.49\textwidth]{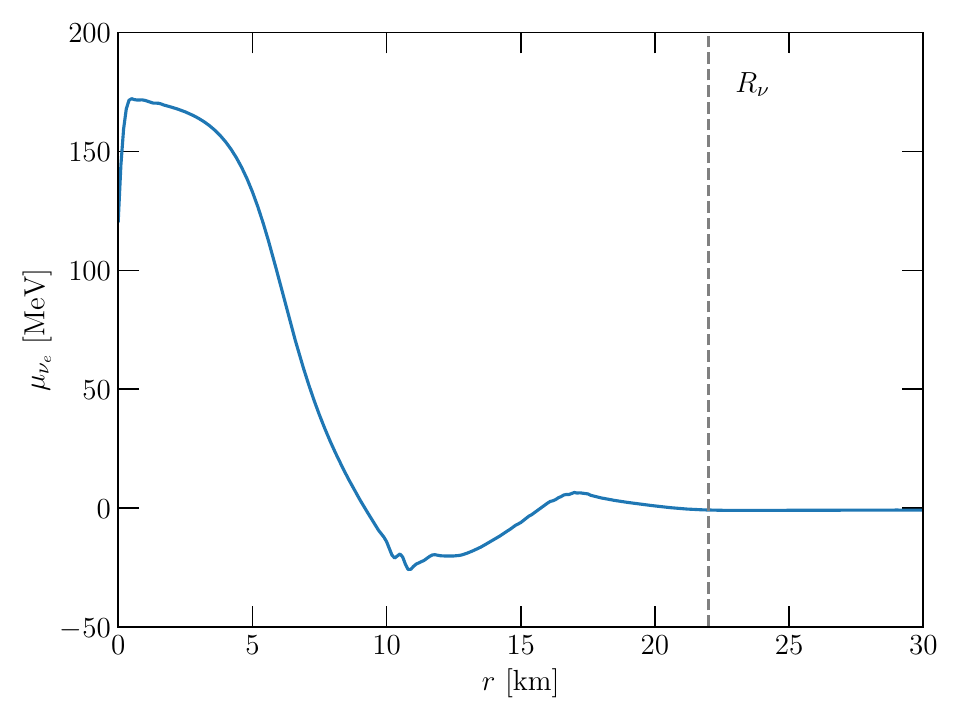}
    \includegraphics[width=0.49\textwidth]{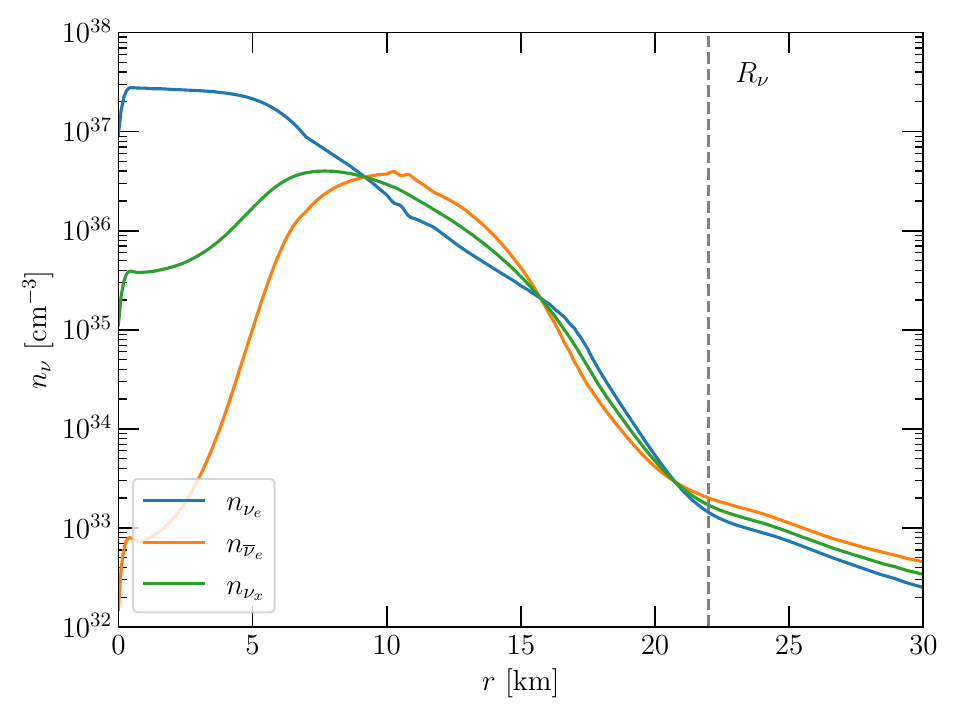}
    
    \caption{Profiles of baryon density $\rho$, temperature $T$, electron neutrino chemical potential $\mu_{\nu_e}$ and number densities of neutrinos and antineutrinos at 1.0~s post core bounce in a CCSN simulation with a $20\,M_\odot$ progenitor and standard physics. The vertical dashed lines indicate the position of the neutrino sphere ($R_\nu$).}
    \label{fig:snsim}
\end{figure*}
%%%%%%%%%%%%%%%%%%%
\acknowledgments
 
We would like to thank Hans-Thomas Janka for providing access to simulation data through the Garching Core Collapse Supernova Archive. The particular simulation whose output was used for our work was done by Robert Bollig and Lorenz H\"udepohl using the \texttt{PROMETHEUS-VERTEX} code. In addition, we thank Gonzalo Herrera, Manibrata Sen and Robert Ziegler for useful discussions. B.D. also thanks Kaladi Babu, Doojin Kim, Deepak Sathyan, Kuver Sinha, Anil Thapa and  Yongchao Zhang for collaboration on related topics, and the organizers of the BCVSPIN 2024 conference in Kathmandu, Nepal, for warm hospitality where part of this work was done. The work of B.D. was partly supported by the US
Department of Energy under grant No. DE-SC0017987. C.V.C. was generously supported by the Arthur B. McDonald Canadian Astroparticle Physics Research Institute, and by Washington University in St. Louis through the Edwin Thompson Jaynes Postdoctoral Fellowship.  A.V.P. thanks Sanjay Reddy for helpful discussions and acknowledges partial support from the U.S. Department of Energy (DOE) under grant DE-FG02-87ER40328 at the University of Minnesota. 

\appendix

\section{Appendix: Simulation Inputs}\label{Appendix A}

For the calculations presented in this paper, we use the output of a CCSN simulation for a $20\,M_\odot$ progenitor with the SFHo equation of state and without muons in the PNS~\cite{jankasim}. We take a snapshot of the baryon density $\rho$, temperature $T$, electron neutrino chemical potential $\mu_{\nu_e}$ and number densities of neutrinos and antineutrinos of all flavors as functions of radius $r$ at 1.0~s post core bounce. These profiles are depicted in Fig.~\ref{fig:snsim}. This particular snapshot is chosen because it has the maximum overall DM luminosity, see Fig.~\ref{fig:emissionvstime}.

%%%%%%%%%%%%%%%%%%%%%%%%%%%%%%%%%%%%%%%%%%%%%%%%%%%%%%%%%%%%%%%%%%
%%%%%%%%%%%%%%%%%%%%%%%%%%%%%%%%%%%%%%%%%%%%%%%%%%%%%%%%%%%%%%%%%%
\bibliography{main}
\end{document}